\documentstyle[12pt,epsf]{article}
\newcommand{\be}{\begin{equation}}
\newcommand{\ee}{\end{equation}}
\newcommand{\ba}{\begin{eqnarray}}
\newcommand{\ea}{\end{eqnarray}}
\newcommand{\bb}{\begin{thebibliography}}
\newcommand{\eb}{\end{thebibliography}}
\newcommand{\ci}[1]{\cite{#1}}
\newcommand{\bi}[1]{\bibitem{#1}}
\newcommand{\lab}[1]{\label{#1}}

\begin{document}
\phantom{.}
\vspace{2cm}
\begin{center}
{\bf {\large{On Possible Study of Quark--Pomeron Coupling
Structure at the COMPASS spectrometer}}} \\

S.V.Goloskokov,
\footnote{Email:  goloskkv@thsun1.jinr.dubna.su}\\
Bogoliubov Laboratory of Theoretical Physics,\\ Joint Institute for
Nuclear Research,\\ Dubna 141980, Moscow region, Russia
\end{center}

\vspace{.7cm}
\begin{abstract}
We analyse the diffractive $Q \bar Q$ production and final jet
kinematics
in polarized deep--inelastic $lp$ scattering at $\sqrt{s}=20 GeV$.
We show that
this reaction can be used in the new spectrometer of the COMPASS
Collaboration at CERN to study the quark--pomeron coupling structure.
\end{abstract}
\newpage
The diffractive events with a large rapidity gap in deep inelastic
lepton--proton scattering
\be
e+p \to e'+p'+X                          \lab{de} \ee
have recently been
investigated  (see, e.g. \ci{gap1,gap2}).  These experiments
have given an excellent tool to test the structure of the pomeron
and its couplings. As a result, the study of the pomeron properties
becomes again popular now.

The  diffractive lepton-proton reactions (\ref{de})  is described
usually in terms of the kinematic variables
\ba
Q^2=-q^2,\;t=r^2, \nonumber
\\  y=\frac{pq}{p_l p},\;x=\frac{Q^2}{2pq},\;
x_p=\frac{q(p-p')}{qp},\;\beta=\frac{x}{x_p},
\ea
where $p_l,p'_l$ and $p, p'$ are the initial and final lepton and
proton momenta, respectively, $q=p_l-p'_l, r=p-p'$ are the virtual
photon and pomeron momenta.

The cross section of this reaction is related
to the diffractive structure function
\be
\frac{d^4\sigma}{dx dQ^2 dx_p dt}=
\frac{4 \pi \alpha^2}{x Q^4}[1-y+\frac{y^2}{2}]
F_2^{D(4)}(x,Q^2,x_p,t),
\lab{f2}
\ee
which is determined by the pomeron contribution and usually
represented at small $x_p$ in the factorized form
\be
 F_2^{D(4)}(x,Q^2,x_p,t)=f(x_p,t) F_2^{P}(\beta,Q^2,t).
 \ee
 Here $f(x_p,t)$ is the pomeron flux factor  and
 $F_2^{P}(\beta,Q^2,t)$ is the pomeron structure function.

The function $f(x_p,t)$ at small $x_p$ behaves as \ci{pom}
\be
f(x_p,t) \propto \frac{1}{x_p^{2\alpha_P(t)-1}},        \lab{flux}
\ee
where  $\alpha_P(t)$ is the pomeron trajectory
\be
\alpha_P(t)=\alpha_P(0)+\alpha' t,\;\;\;\; \alpha'=0.25(GeV)^{-2}.
\ee

The future polarized diffractive experiments at
DESY, CERN and Brookhaven \ci{bu-now-sch} might give the possibility
to
study the spin structure of the pomeron. One of the places to perform
such an experiment is the future detector of the COMPASS
collaboration
at CERN \ci{compass} which will use
the polarized muon beam and fixed polarized hadron target. The
important feature of COMPASS is the possibility to detect the hadron
component of the process within an angle of about 200-250 Mrad.

The question on the value of the spin--flip component of the pomeron
should be very
important for the diffractive scattering of polarized particles. In
the
nonperturbative two-gluon exchange model \ci{la-na} and the BFKL
model
\cite{bfkl} the pomeron couplings have a simple matrix structure (the
standard coupling in what follows):
\be
V^{\mu}_{hh I\hspace{-1.1mm}P}
=\beta_{hh I\hspace{-1.1mm}P}\; \gamma^{\mu}. \lab{pmu}
\ee
In this case,
the spin-flip effects are suppressed as a power of $s$.

It was shown in \cite{golpl,golpr}  that in addition to the
standard
 pomeron vertex (\ref{pmu}) determined by the diagrams where gluons
 interact
with one quark in the hadron  \cite{la-na}, the large-distance
gluon-loop effects (see Fig.1)
should complicate the structures of the pomeron coupling.
Really, if we consider the gluon loop correction of Fig.1a for the
standard pomeron vertex (\ref{pmu}) and the massless quark,
 we obtain in addition to the $\gamma_\mu$ term, new structures
\begin{equation}
\gamma_{\alpha}(/ \hspace{-2.3mm} k+/ \hspace{-2.3mm} r) \gamma^{\mu}
/ \hspace{-2.3mm} k \gamma^{\alpha} \simeq -2[2(/ \hspace{-2.3mm} k+
\frac{/ \hspace{-2.3mm} r}{2}) k^\mu+i \epsilon^{\mu\alpha\beta\rho}
k_\alpha r_\beta \gamma_\rho \gamma_5],
\end{equation}
where $k$ is a quark momentum, $r$ is a momentum transfer.
 The perturbative calculations \cite{golpl} of both graphs, Fig.1,
give the following form for this vertex:
\begin{equation}
V_{qqI\hspace{-1.1mm}P}^{\mu}(k,r)=\gamma^{\mu} u_0+2 M_Q k^{\mu}
u_1+
2 k^{\mu}
/ \hspace{-2.3mm} k u_2 + i u_3 \epsilon^{\mu\alpha\beta\rho}
k_\alpha r_\beta \gamma_\rho \gamma_5+i M_Q u_4
\sigma^{\mu\alpha} r_\alpha,    \label{ver}
\end{equation}
where $M_Q$ is the quark mass. We shall call the form (\ref{ver}) the
spin-dependent pomeron
coupling.  It has been
shown \cite{gol4} that the functions $u_1(r)-u_4(r)$ can reach $20 -
30 \%$ of the
standard pomeron term $\sim \gamma_{\mu}$ for $|r^2| \simeq {\rm
few}~
GeV^2$. Moreover, they result in the spin-flip effect
at the quark-pomeron vertex in contrast with the term $\gamma_\mu$.
So, the loop diagrams lead to a complicated spin structure of the
pomeron couplings. The phenomenological vertex
$V_{qqI\hspace{-1.1mm}P}^{\mu}$
with the $\gamma_\mu$ and $u_1$ terms was proposed in \ci{klen}. The
modification of the standard pomeron vertex (\ref{pmu}) might be
obtained from the instanton contribution  \ci{inst}.

The test of the spin properties of the pomeron coupling can be done
in future polarized experiments.
At small $x_p$, the
contribution where all the energy of the pomeron goes into the $Q
\bar Q$ production \cite{lanj,coll} might be very important. The role
of these contributions in the spin asymmetries of diffractive
two--jet
production
has been studied in \ci{golpr,golall}.
It has been found that the $A_{ll}$ asymmetry in the light quark
production in deep inelastic $l p$ scattering, Fig.2, is dependent on
the pomeron coupling structure.
This asymmetry for cross sections integrated over
the transverse momentum of jet could reach $10-20\%$ \ci{golall}.
The dependence of polarized cross sections and double--spin
longitudinal asymmetry on the transverse momentum of a produced jet
$k_\perp^2$ and their sensitivity to the quark--pomeron coupling
structure have been studied in \ci{golhera}.

In this paper, we analyze the effects of the quark--pomeron
coupling  in the polarized diffractive $e+p \to e'+p'+Q \bar Q$
reaction at the energy
$\sqrt{s}=20GeV$. We estimate the cross section, the longitudinal
double--spin asymmetry $A_{ll}$ and the
kinematics of the final jet to show that these
events can be studied at future spectrometer of the COMPASS
Collaboration \ci{compass}.

The diffractive light  $Q \bar Q$ production in lepton-proton
reaction is determined by the diagram of Fig. 2.
The spin-average cross section can be written in the form
\ci{golhera}
\ba
\sigma(t)=
\frac{d^5 \sigma(^{\rightarrow} _{\Leftarrow})}{dx dy dx_p dt
dk_\perp^2}+
\frac{d^5 \sigma(^{\rightarrow} _{\Rightarrow})}{dx dy dx_p dt
dk_\perp^2}= \nonumber\\
 \frac{3(1-y+y^2/2)\beta_0^4 F(t)^2 [9\sum_{i}e^2_i] \alpha^2}{128
x_p^{2\alpha_{P}(t)} y Q^2 \pi^3} \frac{N(\beta,k_\perp^2,x_p,t)}
{\sqrt{1-4k_\perp^2\beta/Q^2}(k_\perp^2+M_Q^2)^2}. \lab{sigma}
\ea
Here  $\sigma(^{\rightarrow} _{\Rightarrow})$ and
$\sigma(^{\rightarrow}
_{\Leftarrow})$ are the cross sections with parallel and antiparallel
longitudinal polarization of the leptons and protons,
$\beta_0$ is the quark--pomeron coupling, $F(t)$
is the pomeron-proton form factor, $e_i$ are the quark charges.
The leading $x_p$ dependence is extracted in the coefficient of
Eq.(\ref{sigma}) which is determined by the pomeron flux factor
(\ref{flux}). The  trace over the quark loop --$N$ may be decomposed
as follows
\be
 N(\beta,k_\perp^2,t)= N^s(\beta,k_\perp^2,t)+
 \delta N(\beta,k_\perp^2,t).   \lab{nd}
 \ee
Here $N^s$ is the contribution of the standard pomeron vertex
(\ref{pmu}) and $\delta N$ contains the contribution of the
$u_1(r)-u_4(r)$
 terms from (\ref{ver}). For $N^s$ in the case of light
 quarks in the loop and $x_p=0$ we find
\be
N^s(\beta,k_\perp^2,t)=32 [2(1-\beta) k_\perp^2-\beta |t|]|t|.
\lab{ns}
\ee
 The form of $\delta N$  is  more complicated.
We have found it in the $\beta \to 0$ limit. For the massless
quarks only the $u_3$ terms contribute to $\delta N$:
\be
\delta N(k_\perp^2,t) = 32 k_\perp^2 |t| [(k_\perp^4+4
k_\perp^2 |t|+|t|^2) u_3- 4 k_\perp^2-2 |t|] u_3.  \lab{deld}
\ee
Note that $\delta N$ is positive because $u_3 \le 0$. Higher twist
terms
of an order of $M_Q^2/Q^2$ and $|t|/Q^2$ have been dropped in
(\ref{ns},\ref{deld}).

The difference of
the cross section for the supercritical pomeron can be written
in the form
\ba
\Delta \sigma(t)=
\frac{d^5 \sigma(^{\rightarrow} _{\Leftarrow})}{dx dy dx_p dt
dk_\perp^2}-
\frac{d^5 \sigma(^{\rightarrow} _{\Rightarrow})}{dx dy dx_p dt
dk_\perp^2}= \nonumber\\
 \frac{3(2-y)\beta_0^4 F(t)^2 [9\sum_{i}e^2_i] \alpha^2}{128
x_p^{2\alpha_{P}(t)-1} Q^2 \pi^3} \frac{A(\beta,k_\perp^2,x_p,t)}
{\sqrt{1-4k_\perp^2\beta/Q^2}(k_\perp^2+M_Q^2)^2}. \lab{dsigma}
\ea
The notation here is similar to that used in Eqs.
(\ref{sigma}).

The function $A$ is determined by the trace over the quark loop.
It can be written in the $x_p \to 0$ limit as follows:
\be
 A(\beta,k_\perp^2,t)= A^s(\beta,k_\perp^2,t)+
 \delta A(\beta,k_\perp^2,t).   \lab{na}
 \ee
Here $A^s$ is the contribution of the standard pomeron vertex
(\ref{pmu}) and $\delta A$ is determined by the
$u_1(r)-u_4(r)$ terms from (\ref{ver}).

The function $A^s$ for the light quarks looks like
\be
 A^s(\beta,k_\perp^2,t)=
   16 (2 (1-\beta) k_\perp^2 - |t| \beta) |t|.   \lab{ad}
 \ee
We have calculated $\delta A$ in the $\beta \to 0$ limit.
 For the massless quarks we have
\be
 \delta A(\beta,k_\perp^2,t)= - 16 (3 k_\perp^2 + 2 |t|) k_\perp^2
 |t|
    u_3.
      \lab{dad} \ee
The leading twist terms have been calculated here as previously.

It can be seen that $\sigma$ has a more singular behaviour  than
$\delta
\sigma$ as $x_p \to 0$. This is determined by the fact that the
leading
term in $\delta \sigma$ is proportional to
$\epsilon^{\mu\nu\alpha\beta}r_\beta...\propto x_p p$. The same is
true for
the lepton part of the diagram of Fig.1. As a result, the
additional term $y x_p$ appears in $\delta \sigma$.

 We calculate the cross section
 integrated over momentum transfer because it is difficult to detect
 the recoil proton in COMPASS detector
\be
\sigma[\Delta \sigma]=\int_{t_m}^{0} dt
\sigma(t)[\Delta \sigma(t)],\;\;\;|t_m|=7(GeV)^2. \lab{intsi}
\ee
The exponential form of the proton form factor $F(t)=e^{bt}$   with
$b=1.9(GeV)^{-2}$ has been used.

As an example, we calculate the cross sections and
asymmetry for  $\beta=0.175, y=0.7, x_p=0.1 \;{\rm
and}\; Q^2=5GeV^2$. The results for the
 cross section of the light quark production in diffractive deep
inelastic scattering for the pomeron with the pomeron intercept
$\alpha_{P}(0)= 1.1$ are shown in Fig. 3 for the standard
and
spin-dependent pomeron couplings.   The shape of  both the curves is
very similar and for the
spin-dependent pomeron coupling the cross section is almost twice
that for the standard pomeron coupling.

The longitudinal double spin asymmetry is determined by the relation
\be
A_{ll}=
\frac{\Delta \sigma}{\sigma}=\frac{
\sigma(^{\rightarrow} _{\Leftarrow})-\sigma(^{\rightarrow}
_{\Rightarrow})}
{\sigma(^{\rightarrow} _{\Rightarrow})+\sigma(^{\rightarrow}
_{\Leftarrow})}, \lab{asydef}
\ee
The asymmetry of the diffractive light $Q \bar Q$ production is shown
in Fig. 4.
It can be seen from the cross section (\ref{dsigma},\ref{sigma}) that
the asymmetry for the standard quark--pomeron vertex is very simple
in
form
\be
A_{ll}=\frac{y x_p (2-y)}{2-2y+y^2}.
\ee
There is no any $k_\perp$ and $\beta$ dependence here. For the
spin--dependent pomeron coupling the asymmetry is more  complicated
because of different contributions to $\delta A$ and
$\delta N$  proportional to $k^2_{\perp}$.
In this case the $A_{ll}$ asymmetry is smaller than for the standard
 pomeron vertex. Thus, the
$A_{ll}$ asymmetry can be used to test the quark-pomeron coupling
structure.

Let us estimate now the kinematics of jet events.  The jet momenta
are:
\be
j_1=q-k, \;\;j_2=r+k. \lab{jet}
\ee
The photon momentum can be written in the center-mass
system in the form
\be
q=\left( y \sqrt{s},\frac{-Q^2}{\sqrt{s} }, \vec q_{\perp}
\right),\lab{m1}
\;\;
|\vec q_{\perp}|=\sqrt{(1-y) Q^2}.
\ee
The transverse momentum $r$ can be written as follows
\be
r=\left( \frac{-|t|}{\sqrt{s}}, x_p \sqrt{s}, \vec r_{\perp} \right),
\;\;
|\vec r_{\perp}|=\sqrt{(1-x_p) |t|}.\lab{m2}
\ee
From the mass-shell conditions for jet momenta $j_1^2=j_2^2=M_Q^2$
the quark momentum $k$ has been found to be
\be
k \simeq \left(\frac{(\vec r_{\perp}+\vec
k_{\perp})^2+M_Q^2}{\sqrt{s}
x_p},-\frac{y Q^2+(\vec q_{\perp}-\vec
k_{\perp})^2+M_Q^2}{\sqrt{s} y}, \vec k_{\perp}\right).\lab{m3}
\ee
In (\ref{m1}-\ref{m3}) the light-cone variables have been used.

The jet momenta and its angles in the rest system of the initial
proton can be
expressed in terms of (\ref{m1}-\ref{m3})
\ba P_{J1} \simeq \frac{y x_p s -k_{\perp}^2-M_Q^2}{2 x_p m},\;
\sin{\left(\frac{\theta_{J1}}{2}\right)} \simeq \frac{m \sqrt{(\vec
k_{\perp}-\vec
q_{\perp})^2}}{y s};\\ \nonumber
P_{J2} \simeq \frac{ m^2+M_Q^2+k_{\perp}^2}{2 x_p m},\;
\sin{\left(\frac{\theta_{J2}}{2}\right)}  \simeq \frac{m
x_p}{\sqrt{m^2+M_Q^2+k_{\perp}^2}}.
\ea
Here m is the proton mass. The invariant mass of a produced system is
\be
M_{2Jet}^2=x_p y s.
\ee
The momenta and jet angles for $\sqrt{s}=20GeV, x_p=0.1, y=0.7 \;{\rm
and}\;
Q^2=5GeV^2$ are shown in Figs 5,6 for the azimuth angle between the
lepton scattering plane and $k_{\perp}$ is equal to 90 degree.
It is seen that both jets can be detected by the COMPASS detector
whose angular acceptance is about 200-250Mrad.\\[0.5cm]

Thus, we have found that the  structure of the quark--pomeron
coupling should modify the
spin average and  spin--dependent cross section.
The spin--dependent form of $V_{qqP}$ almost twice increases the
cross section.
However, the shape of the cross
sections is very similar for the standard and spin--dependent pomeron
 vertices. The $A_{ll}$ asymmetry is more
convenient to test the pomeron coupling
structure. The asymmetry is free from normalization
factors and is sensitive to the dynamics of pomeron interaction.
We have found a well-defined prediction for $A_{ll}$ for the standard
pomeron vertex. This conclusion is similar to the results of
\ci{golpr} where
the single-spin asymmetry in the diffractive $Q \bar Q$ production
has been studied.

The predicted cross sections are not small for the experimental
investigation of this reaction. Our analysis of jet kinematics shows
that they might be detectable by the COMPASS  spectrometer. There is
no
possibility to detect the final proton. However, the analysis of the
diffractive events similar to that  done in HERA
experiments \ci{gap1,gap2} can be performed in this case, too.

We can conclude that the study of the longitudinal double
spin asymmetry and the cross section of the diffractive deep
inelastic
scattering at the new spectrometer of the COMPASS
Collaboration at CERN can give important information about the
complicated spin
structure or the pomeron coupling.\\[0.5cm]

The author expresses his deep gratitude to  A.V.Efremov, M.Finger,
G.Mallot, P.Kroll, W.-D.Nowak, I.A.Savin, A.Sch\"afer and
O.V.Teryaev for fruitful discussions.

\newpage

  \vspace*{-.5cm}
\epsfxsize=11cm
\centerline{\epsfbox{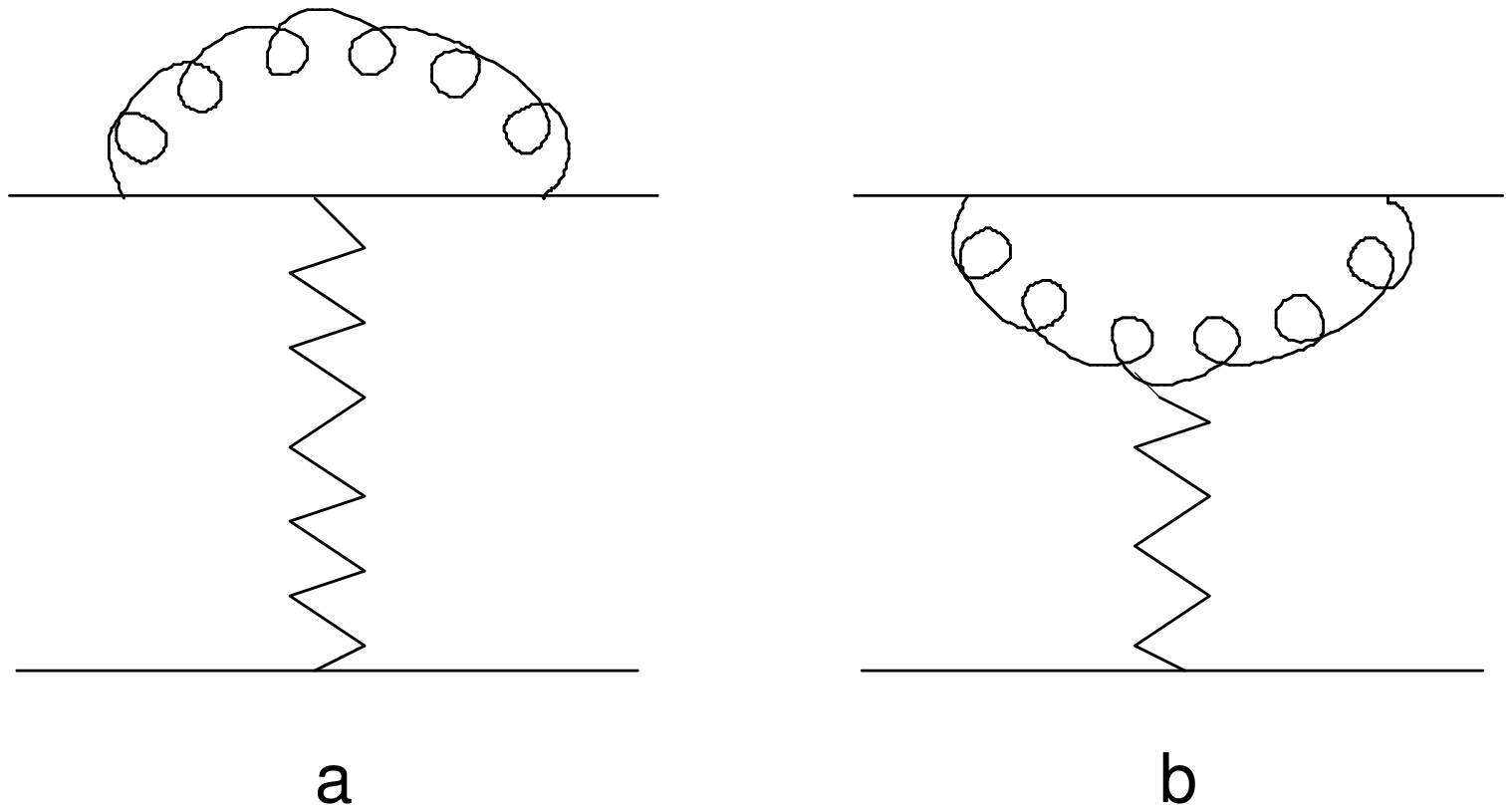}}
  \vspace*{-7.9cm}
\begin{center}
Fig.1 Gluon-loop contribution to the quark-pomeron coupling.
Broken line -the pomeron exchange.
\end{center}
\vspace{1cm}
\samepage
  \vspace*{.8cm}
      \hspace*{3.5cm}
\mbox{
   \epsfxsize=15cm
   \epsffile{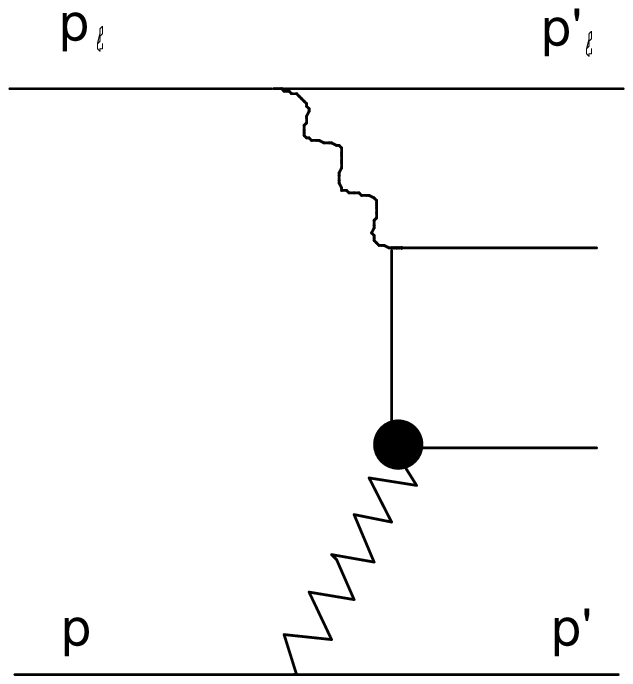}
}
  \vspace*{-12.4cm}
\begin{center}
Fig.2~Diffractive $Q \bar Q$   production in deep
inelastic scattering
\end{center}

\newpage
  \vspace*{-2.5cm}
\epsfxsize=14cm
\centerline{\epsfbox{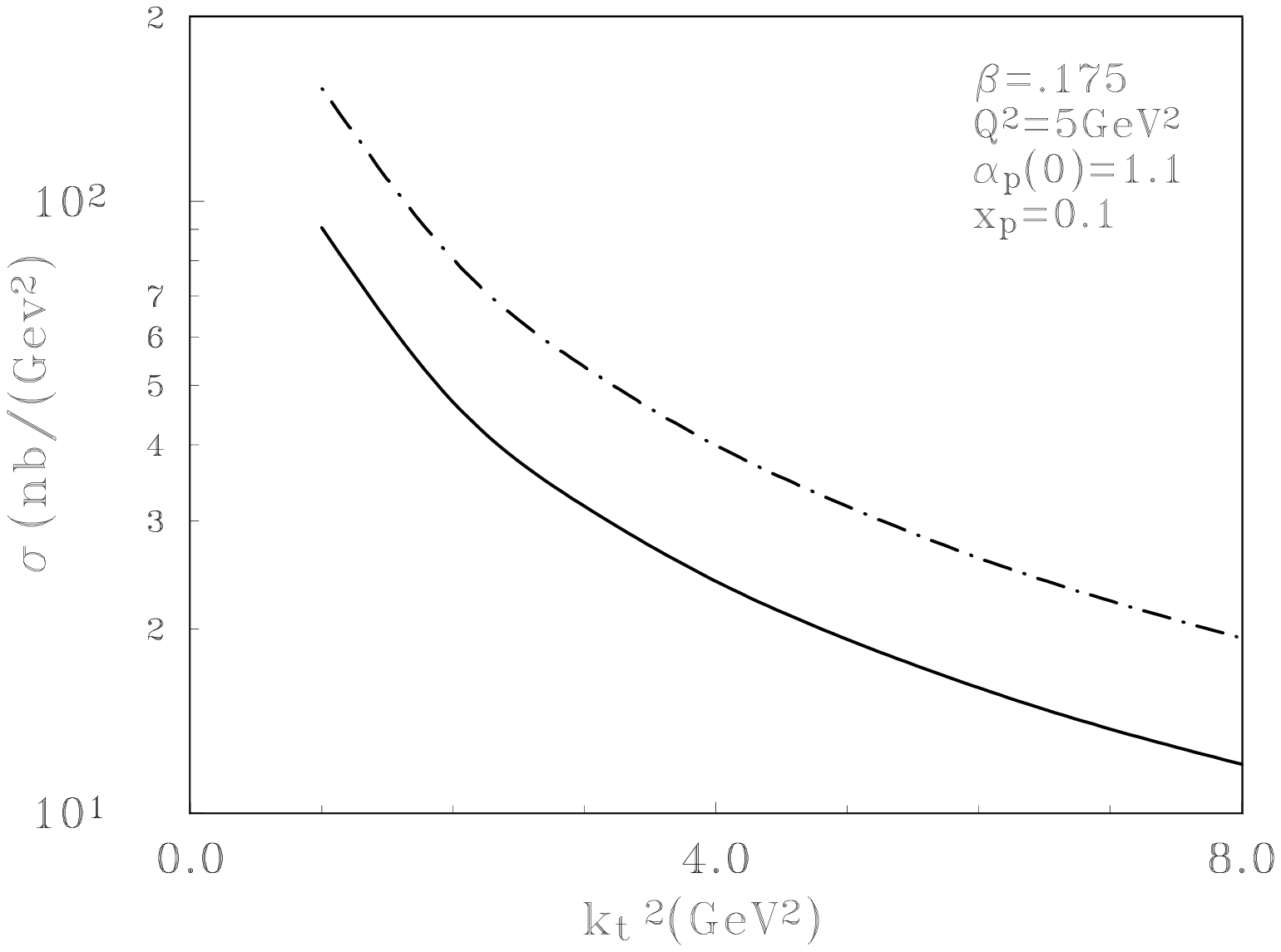}}
  \vspace*{-.1cm}
\begin{center}
Fig.1 ~$k^2_{\perp}$-- dependence of cross-sections at
$\sqrt{s}=20(GeV)$.
Solid line -for the standard vertex;
dot-dashed line -for the spin-dependent quark-pomeron
vertex.\end{center}

  \vspace*{-.6cm}
\epsfxsize=14cm
\centerline{\epsfbox{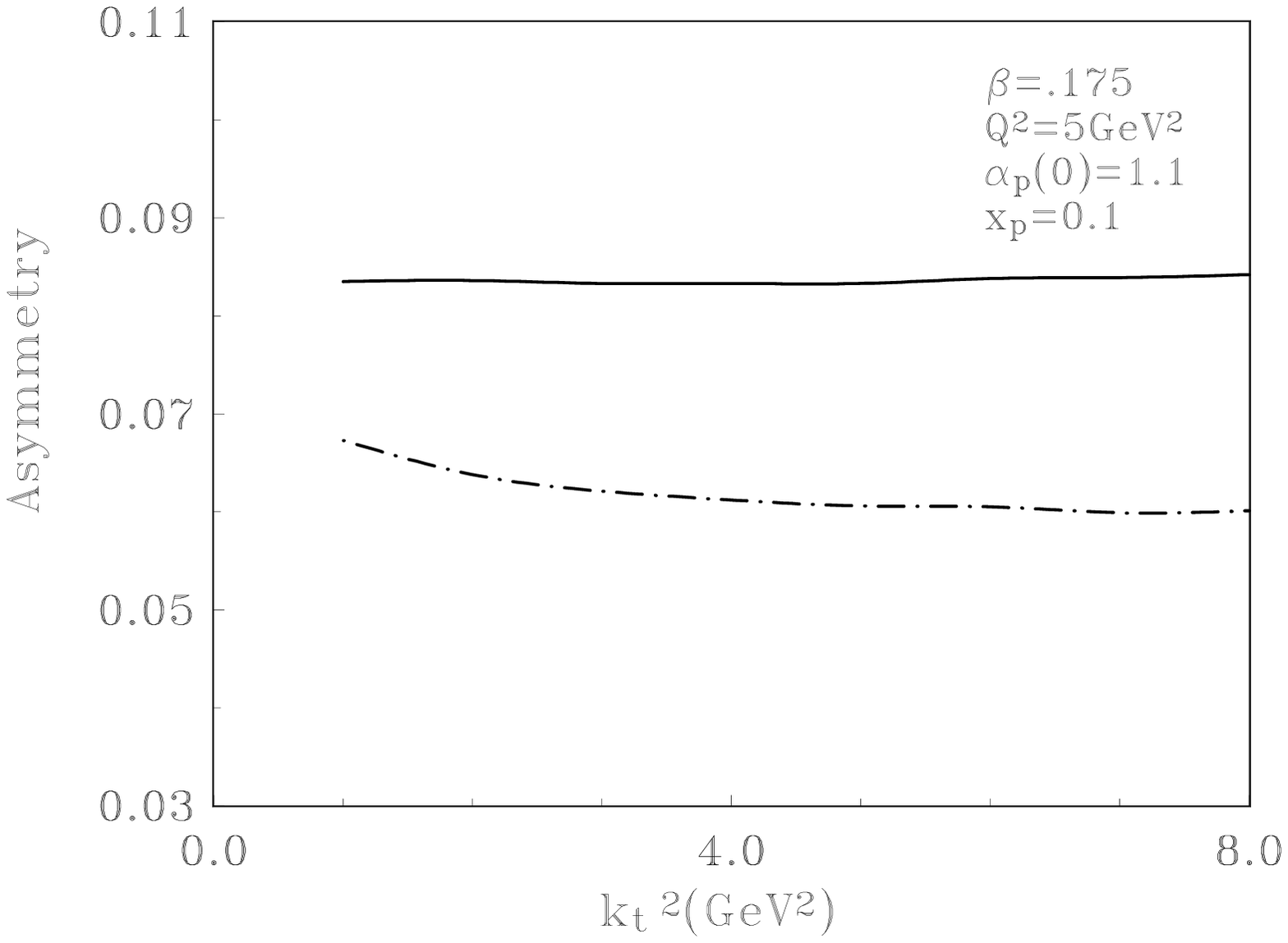}}
  \vspace*{.3cm}
\begin{center}
Fig.2 ~$k^2_{\perp}$-- dependence of $A_{ll}$ asymmetry at
$\sqrt{s}=20(GeV)$.
Solid line -for the standard vertex;
dot-dashed line -for the spin-dependent quark-pomeron vertex.

\end{center}
\newpage
  \vspace*{-2.5cm}
\epsfxsize=14cm
\centerline{\epsfbox{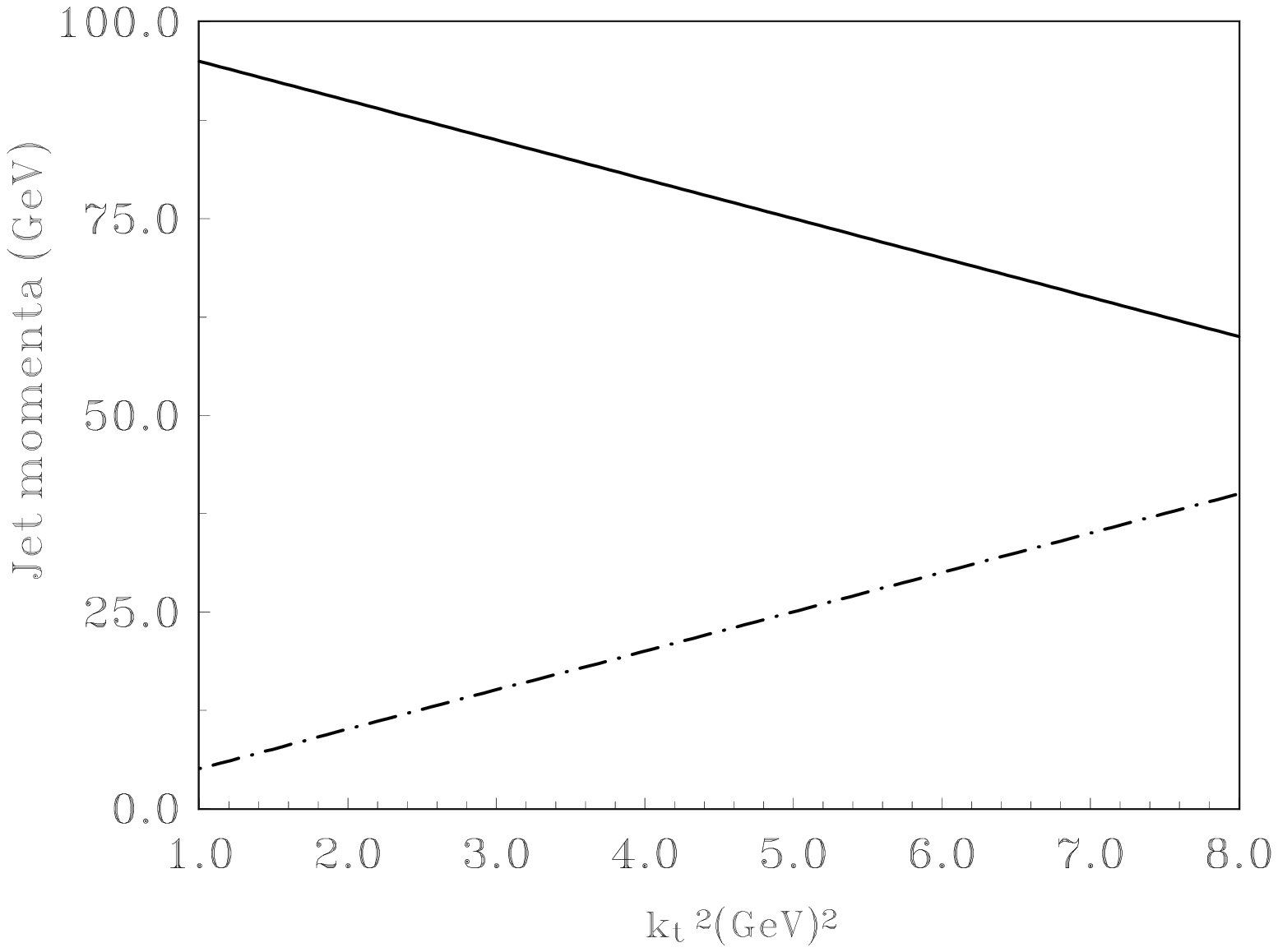}}
  \vspace*{-.1cm}
\begin{center}
Fig.5 ~$k^2_{\perp}$-- dependence of jet momenta.
Solid and
dot-dashed line -for jet1 and jet2 respectively.\end{center}

  \vspace*{-.6cm}
\epsfxsize=14cm
\centerline{\epsfbox{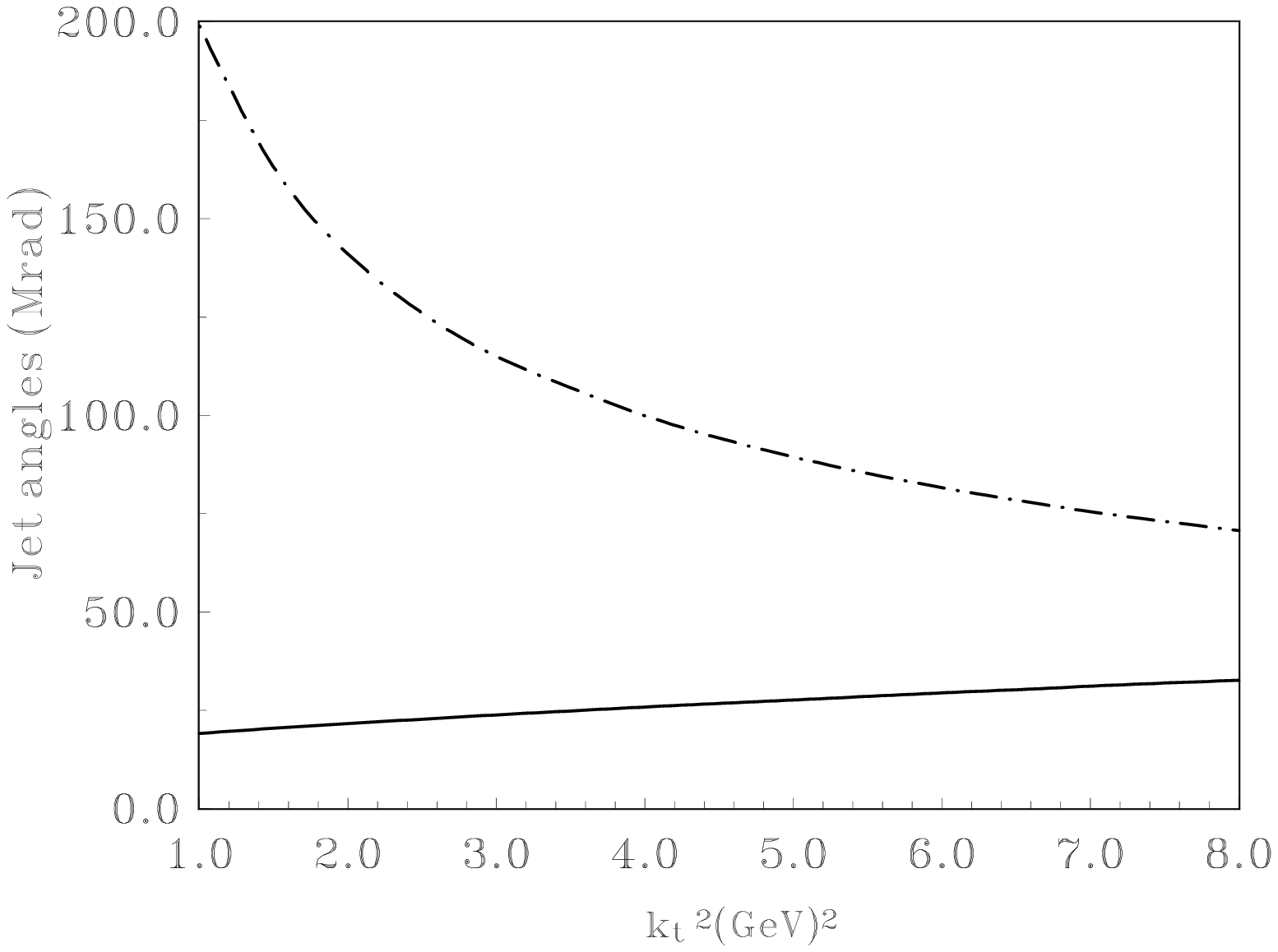}}
  \vspace*{.3cm}
\begin{center}
Fig.6 ~$k^2_{\perp}$-- dependence of jet angles.
Solid and
dot-dashed line -for jet1 and jet2 respectively.

\end{center}
\end{document}